\documentclass[twocolumn,aps,superscriptaddress,prl]{revtex4}
\usepackage{graphicx}
\usepackage{xcolor}
\usepackage{times}
\usepackage{amsmath}
\usepackage{amssymb}
\usepackage{units}
\usepackage{stmaryrd} 
\DeclareGraphicsExtensions{.pdf,.png,.eps,.jpg}

\newcommand{\DUV}{DFT$\,+U$$\,+V$}

\usepackage{xcolor}

\usepackage{ulem}

\begin{document}
\preprint{0}

\title{Light-induced renormalization of the Dirac quasiparticles in the nodal-line semimetal ZrSiSe}

\author{G. Gatti$^\star$} 
\affiliation{Ecole Polytechnique F\'ed\'erale de Lausanne (EPFL), CH-1015 Lausanne, Switzerland} 
\affiliation{Lausanne Centre for Ultrafast Science (LACUS), Ecole Polytechnique F\'ed\'erale de Lausanne (EPFL), CH-1015 Lausanne, Switzerland} 

\author{A. Crepaldi$^\star$} \email[E-mail address: ]{alberto.crepaldi@epfl.ch}
\affiliation{Ecole Polytechnique F\'ed\'erale de Lausanne (EPFL), CH-1015 Lausanne, Switzerland}
\affiliation{Lausanne Centre for Ultrafast Science (LACUS), Ecole Polytechnique F\'ed\'erale de Lausanne (EPFL), CH-1015 Lausanne, Switzerland} 

\author{M. Puppin} 
\affiliation{Lausanne Centre for Ultrafast Science (LACUS), Ecole Polytechnique F\'ed\'erale de Lausanne (EPFL), CH-1015 Lausanne, Switzerland} 
\affiliation{Laboratory of Ultrafast Spectroscopy, ISIC, Ecole Polytechnique F\'ed\'erale de Lausanne (EPFL), CH-1015 Lausanne, Switzerland} 

\author{N. Tancogne -Dejean} 
\affiliation{Max Planck Institute for the Structure and Dynamics of Matter and Center for Free-Electron Laser Science, Luruper Chaussee 149, 22761 Hamburg, Germany}

\author{L. Xian} 
\affiliation{Max Planck Institute for the Structure and Dynamics of Matter and Center for Free-Electron Laser Science, Luruper Chaussee 149, 22761 Hamburg, Germany}

\author{S. Roth} 
\affiliation{Ecole Polytechnique F\'ed\'erale de Lausanne (EPFL), CH-1015 Lausanne, Switzerland} 
\affiliation{Lausanne Centre for Ultrafast Science (LACUS), Ecole Polytechnique F\'ed\'erale de Lausanne (EPFL), CH-1015 Lausanne, Switzerland} 

\author{S. Polishchuk} 
\affiliation{Lausanne Centre for Ultrafast Science (LACUS), Ecole Polytechnique F\'ed\'erale de Lausanne (EPFL), CH-1015 Lausanne, Switzerland} 
\affiliation{Laboratory of Ultrafast Spectroscopy, ISIC, Ecole Polytechnique F\'ed\'erale de Lausanne (EPFL), CH-1015 Lausanne, Switzerland} 

\author{Ph. Bugnon}
\affiliation{Ecole Polytechnique F\'ed\'erale de Lausanne (EPFL), CH-1015 Lausanne, Switzerland}

\author{A. Magrez}
\affiliation{Ecole Polytechnique F\'ed\'erale de Lausanne (EPFL), CH-1015 Lausanne, Switzerland}

\author{H. Berger}
\affiliation{Ecole Polytechnique F\'ed\'erale de Lausanne (EPFL), CH-1015 Lausanne, Switzerland}
 
\author{F. Frassetto} 
\affiliation{National Research Council of Italy - Institute of Photonics and Nanotechnologies (CNR-IFN), via Trasea 7, 35131 Padova, Italy}

\author{L. Poletto} 
\affiliation{National Research Council of Italy - Institute of Photonics and Nanotechnologies (CNR-IFN), via Trasea 7, 35131 Padova, Italy}

\author{L. Moreschini}
\affiliation{Advanced Light Source, Lawrence Berkeley National Laboratory, Berkeley, California 94720, USA}

\author{S. Moser}
\affiliation{Advanced Light Source, Lawrence Berkeley National Laboratory, Berkeley, California 94720, USA}
\affiliation{Physikalisches Institut and W\"urzburg-Dresden Cluster of Excellence ct.qmat, Universit\"at W\"urzburg, 97074 W\"urzburg, Germany}

\author{A. Bostwick}
\affiliation{Advanced Light Source, Lawrence Berkeley National Laboratory, Berkeley, California 94720, USA}

\author{E. Rotenberg}
\affiliation{Advanced Light Source, Lawrence Berkeley National Laboratory, Berkeley, California 94720, USA}

\author{A. Rubio} 
\affiliation{Max Planck Institute for the Structure and Dynamics of Matter and Center for Free-Electron Laser Science, Luruper Chaussee 149, 22761 Hamburg, Germany}
\affiliation{Nano-Bio Spectroscopy Group, Departamento de Fisica de Materiales, Universidad del Pa\'is Vasco, 20018 San Sebastian, Spain}

\author{M. Chergui} 
\affiliation{Lausanne Centre for Ultrafast Science (LACUS), Ecole Polytechnique F\'ed\'erale de Lausanne (EPFL), CH-1015 Lausanne, Switzerland} 
\affiliation{Laboratory of Ultrafast Spectroscopy, ISIC, Ecole Polytechnique F\'ed\'erale de Lausanne (EPFL), CH-1015 Lausanne, Switzerland} 

\author{M. Grioni} 
\affiliation{Ecole Polytechnique F\'ed\'erale de Lausanne (EPFL), CH-1015 Lausanne, Switzerland} 
\affiliation{Lausanne Centre for Ultrafast Science (LACUS), Ecole Polytechnique F\'ed\'erale de Lausanne (EPFL), CH-1015 Lausanne, Switzerland}

\date{\today}

\begin{abstract}

In nodal-line semimetals linearly dispersing states form Dirac loops in the reciprocal space, with high degree of electron-hole symmetry and almost-vanishing density of states near the Fermi level. The result is reduced electronic screening and enhanced correlations between Dirac quasiparticles. Here we investigate the electronic structure of ZrSiSe, by combining time- and angle-resolved photoelectron spectroscopy with \textit{ab initio} density functional theory (DFT) complemented by an extended Hubbard model (\DUV). We show that electronic correlations are reduced on an ultrashort timescale by optical excitation of high-energy electrons-hole pairs, which transiently screen the Coulomb interaction. Our findings demonstrate an all-optical method for engineering the band structure of a quantum material. 

\end{abstract}

\maketitle


The application of topological concepts to condensed matter, which is central to the description of the quantum spin Hall effect \cite{Barnevig_Science_06} and topological insulators \cite{Hasan_RMP_2010}, has been fruitful in the classification of gapless topological phases \cite{Wan_Vishwanath_PRB_11, Burkov_PRB_11, Young_PRL_12}.  Dirac and Weyl fermions emerge as low energy excitations, characterized by a discrete number of symmetry-protected nodes  \cite{Armitage_RMP_18}. The nodes are responsible for magneto-transport properties that are unknown in topologically trivial compounds \cite{Burkov_JPCM_15}. The nodes also exhibit vanishing density of states, which alter the screening of the Coulomb interaction, thus requiring corrections to the Fermi liquid model. Logarithmic corrections play an important role in two-dimensional (2D) compensated semimetals such as graphite \cite{Gonzales_PRB_99} and graphene \cite{Ando_JPSJ_06}. In the case of Dirac and Weyl particles, the reduced screening enhances correlations, affects charge transport \cite{Hosur_PRL_12} and makes the system prone to instabilities towards new ordered phases \cite{Roy_PRB_17, Wang_PRB_13, Yang_NP_14}.

The long-range Coulomb interaction has been theoretically discussed in nodal-line semimetals (NLSMs) \cite{Huh_PRB_16}, where the intersection of linearly dispersing states forms 1D trajectories in reciprocal space \cite{Burkov_PRB_11, Kim_PRL_15, Fang_CPB_16}.  It has been shown that in the low-doping regime of NLSMs, even if the Dirac lines are moved away from the Fermi level, weak screening affects the transport properties \cite{Syzranov_PRB_17}. Similarly to the Dirac and Weyl semimetals, correlations can drive the materials to symmetry-breaking ground states in the bulk  \cite{Roy_PRB_17} and at the surface \cite{Kim_PRL_15}.

NLSMs are realized in several families of square-net materials \cite{Schoop_Review_19}.  ZrXY (X= Si, Sn, Ge; Y = S, Se, Te) is among the most studied, owing to the great flexibility of the chemical composition, and the high crystal quality which allows the quantum limit in transport to be reached at relatively low magnetic field. The appearance of oscillation in the magnetic response \cite{Hu_PRL_117} and in the resistivity \cite{Ali_SA_16, Wang_AEM_16,Singha_PNAS_17}, and the Berry phases extracted from these experiments, bear fingerprints of the non-trivial topology \cite{Ali_SA_16}.  Interestingly, magneto-transport measurements under high magnetic field have revealed an enhancement of the effective mass, which is interpreted as a signature of electronic correlations \cite{Pezzini_NP_17}. These findings  are supported by theoretical models predicting a large excitonic instability in both ZrSiS \cite{Rudenko_PRL_18} and in ZrSiSe \cite{Scherer_PRB_18}, as a consequence of the reduced screening, combined with a large degree of electron-hole symmetry and with a finite density of itinerant charge carriers \cite{Rudenko_PRL_18}. 

Several angle-resolved photoelectron spectroscopy (ARPES) studies have addressed the band structure of ZrSiSe  \cite{Schoop_NatCom_16, Neupane_PRB_16, Chen_PRB_17, Topp_PRX_17, Nakamura_PRB_19}, but no experimental evidence of band renormalization has been reported so far. In this letter, we show that  electronic correlations in ZrSiSe can be modified at the ultrashort timescale by an optical perturbation. The intense laser pulse creates electrons and holes far from the Dirac nodes, which efficiently screen the Coulomb interaction, leading to a renormalization of the Dirac quasiparticles (QP) dispersion probed by time-resolved ARPES (tr-ARPES).  This interpretation is supported by \textit{ab initio} DFT calculations complemented by an extended Hubbard model (\DUV) \cite{intersite}. 

\begin{figure}[bb!]
  \centering   \includegraphics[width = 0.5 \textwidth]{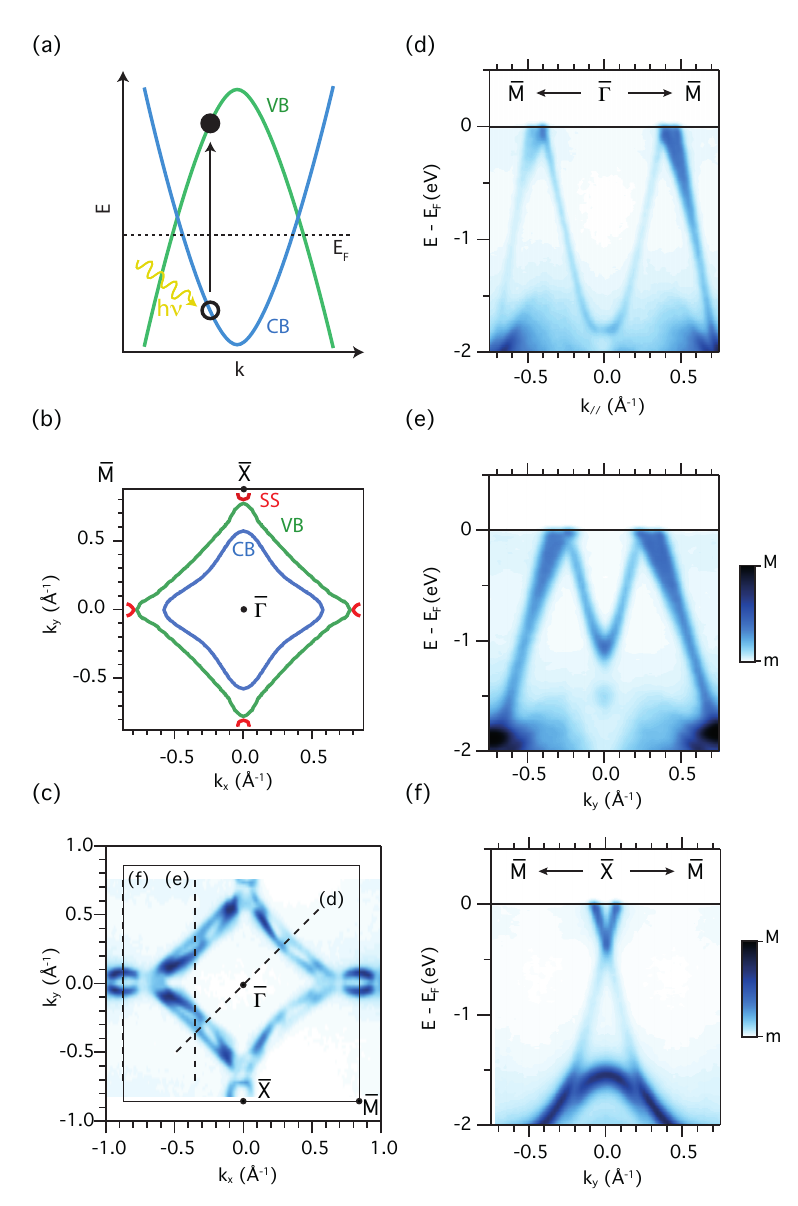}
  \caption[didascalia]{  (c) Sketch of the linearly dispersing Dirac states. Under intense laser light, electrons and holes are injected in the band structure far from the Dirac nodes, and can efficiently screen the Coulomb interaction. Calculated (b) and measured (c) Fermi surface of ZrSiSe. It consists of two diamond-shaped lines centered at $\mathrm{\overline{\Gamma}}$, which originate from the valence band (VB, green) and the conduction band (CB, blue).  An additional surface state (SS, red) is identified around $\mathrm{\overline{X}}$, at the boundary of the surface Brillouin zone  [$\mathrm{\overline{\Gamma} \overline{M} = 1.2}$ \AA $^{-1}$, $\mathrm{\overline{\Gamma} \overline{X} = 0.85}$ \AA $^{-1}$]. 
  (d) - (f) Experimental band dispersion, measured at $143$\, eV photon energy, along the black lines in panel (c). The two linearly dispersing bulk state, shown along the $\mathrm{\overline{\Gamma} - \overline{M}}$ direction in (d), form the Dirac loop, with nodes above the Fermi level. Panels (e) and (f), shown as a reference for the tr-ARPES data in Fig.\,2,  illustrate the dispersion of the bulk (e) and surface (f) states near $\mathrm{E_F}$.
      }
  \label{fig:ARPES}
\end{figure}


High quality single crystals of ZrSiSe grown by vapor transport were cleaved  \textit{in situ} under ultra-high vacuum (UHV) conditions. We measured the band structure by ARPES at the MAESTRO beamline 7.0.2 of the Advanced Light Source in Berkeley, with overall energy and momentum resolutions equal to 0.01 \AA$^{-1}$ and 30 meV. We performed measurements at the tr-ARPES endstation \cite{Crepaldi_Chimia_17} of Harmonium \cite{Ojeda_sd_16} at EPFL in Lausanne. We used s-polarized light at 36.9 eV photon energy, corresponding to the 23rd harmonic generated in argon. The optical excitation was s-polarized and centered at 1.6 eV (780 nm), with an absorbed fluence of 0.38 $mJ/cm^2$ estimated using optical properties from Refs. \cite{Allah_PRB_19,Schilling_PRL_17}. The temporal overlap between pump and probe was determined by the observation of the laser assisted photoelectric effect (LAPE) for p-polarized optical excitation. All measurements were performed at 80 K.

The band-structure calculations were performed on a fully relaxed slab composed of 5 layers of ZrSiSe. We evaluated the on-site $U$ and inter-site $V$ Coulomb terms  \textit{ab initio} and self-consistently using the recently developed extended ACBN0 functional~\cite{intersite} using the Octopus code~\cite{andrade_real-space_2015}. The Hubbard $U$ was evaluated for the $d$ orbitals of Zr, and we restricted ourselves to the nearest-neighbor interaction for the inter-site $V$. We employed fully relativistic pseudopotentials with a $15\times 15 \,\mathbf{k}$-point grid and a grid spacing of 0.158 \AA$^{-1}$ to sample the two-dimensional Brillouin zone. More details about geometry optimization and a comparison with DFT$+U$ and hybrid functionals can be found in the supplemental material~\cite{Supp_ZrSiSe, Weber:2017hq, Johannsen:2013gq, PhysRevB.46.13592, kresse93ab, blochl94, pbe, dftd2, heyd2003hybrid, PhysRevB.97.075140, Agapito_PRX, Implementation_DFTU}.


The band structure of ZrSiSe hosts two kinds of nodal lines; one is protected by the non-symmorphic symmetry and is located far from the Fermi level, $\mathrm{E_F}$, along the bulk $\mathrm{X - R}$ direction \cite{Schoop_NatCom_16, Topp_PRX_17}. The second, more relevant for the screening and the transport properties, is formed by the crossing of the  conduction band (CB) and valence band (VB), as schematized in Fig.\,1(a). Near $\mathrm{E_F}$, the bulk bands are approximately linear, with a small density of states, which is responsible for the relatively poor screening of the long-range Coulomb interaction. The goal of our study is to transiently enhance the screening by optically exciting high-energy electrons-hole pairs far from the Dirac nodes.


The ARPES data of Fig\,1 provide a background for the tr-ARPES experiment. Figure\,1(b) and (c) compare the theoretical and experimental Fermi surface, which consists of two diamond-shaped contours, corresponding to CB (inner) and VB (outer). A surface state (SS) forms short arcs around the $\mathrm{\overline{X}}$ point of the surface Brillouin zone (SBZ) [red line in Fig. 1 (b)] \cite{Schoop_NatCom_16}. Dashed black lines in Fig.\,1(c) indicate the cuts shown in Fig.\,1(d)-(f). Figure\,1(d) displays the bands along the $\mathrm{\overline{\Gamma} - \overline{M}}$ direction, where VB and CB disperse linearly over a broad energy range, in good agreement with the literature \cite{Hosen_PRB_17}.  Figure\,1(e) and (f) show cuts parallel to the $\mathrm{\overline{M} - \overline{X} - \overline{M}}$ direction, intercepting the bulk and the surface states, respectively. 


\begin{figure}[]
  \centering   \includegraphics[width = 0.5 \textwidth]{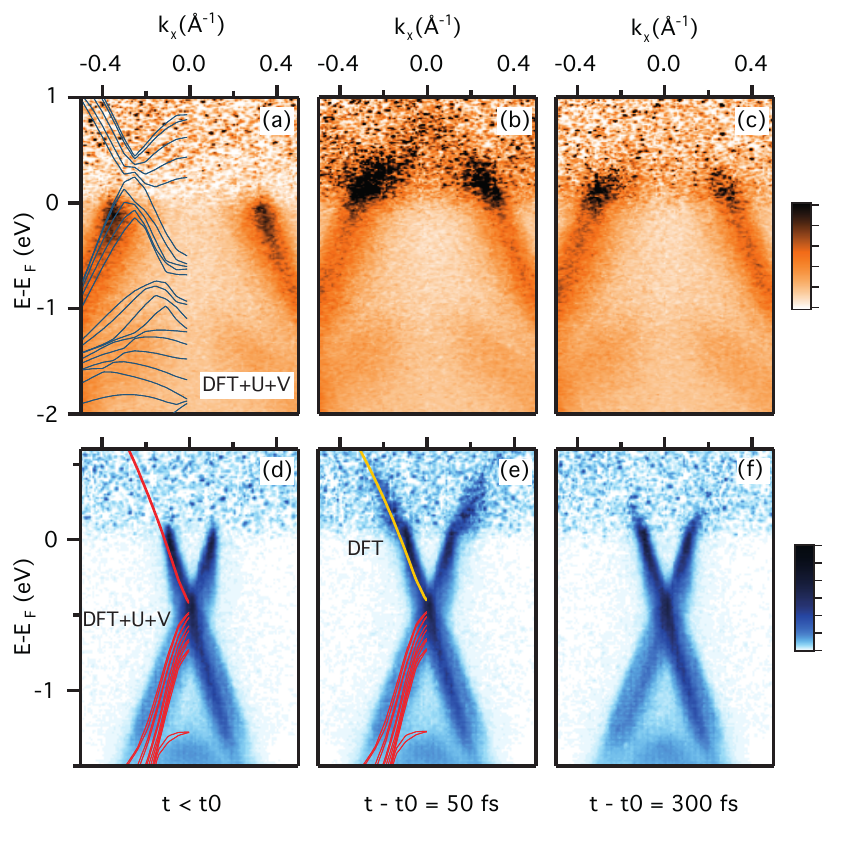}
  \caption[didascalia]{(a)-(c) Band dispersion of the bulk bands, taken along the path (e) indicated in Fig. 1 (c), for different delay times before (a), 50 fs after (b) and 300 fs after (c) the arrival of the optical excitation. Blue  lines indicate the \textit{ab initio} calculations for U = 1.47 eV and V = 0.33 eV.
  (d)-(f) Similar temporal evolution of the band structure along $\mathrm{\overline{M} - \overline{X} - \overline{M}}$ [direction (f) in Fig. 1(c)], focusing on the dispersion of the surface state at different delay times. Red and yellow  lines are the DFT$\,+\,U\,+\,V$ and DFT calculations, respectively.  In both the bulk and the surface states, a change in the band dispersion during optical excitation (50 fs) illustrates the light-induced renormalization of the band dispersion. The intensity at each binding energy is normalized to the same value integrated over the entire momentum window in order to increase the visibility of the signal above $\mathrm{E_F}$.
   }
  \label{fig:tr_ARPES}
\end{figure}


Figure\,2 illustrates the main result of our study. The tr-ARPES data measured at various delay times provide experimental evidence for the light-induced renormalization of the Dirac QP. Figure\,2(a) shows the dispersion of the occupied bulk bands  along the same direction of Fig.\,1(d), for a photon energy of 36.9 eV. Due to the different matrix elements, the intensity of the inner CB state is strongly suppressed. We find that standard DFT calculation fails to reproduce the band dispersion, in particular the second VB state which reaches its maximum 1 eV below $\mathrm{E_F}$. The agreement between theory and experiment is improved by DFT $+U$, with $U=5$\, eV, but this value appears unphysically large, and cannot be reproduced by \textit{ab initio} methods  (see the supplemental material for a detailed comparison \cite{Supp_ZrSiSe}). Due to the partially delocalized character of correlations, the Coulomb interaction is better accounted for by an extended Hubbard model, with $U$ = 1.47 eV and $V$ = 0.33 eV. The  blue lines in Fig.\,2(a) show the corresponding calculated band structure. We stress that $U$ and $V$ are computed fully \textit{ab initio} and they are not free parameters. These moderate values are compatible with several recent observations, for the similar compound ZrSiS: mass enhancement under intense magnetic field \cite{Pezzini_NP_17}, large Land\'e factor in the magnetic response \cite{Hu_PRB_17}, reduced screening of excitons in the low energy optical conductivity \cite{Schilling_PRL_17}.

Figure\,2(b) shows that 50 fs after optical excitation, electrons populate the bulk bands above $\mathrm{E_F}$, and their dispersion seems to deviate from the theoretical predictions. The broadening of the bulk bands, which reflects their intrinsic $k_z$ dispersion, hampers a quantitative analysis of this effect. Therefore, we turn our attention to the sharper surface state, whose unperturbed dispersion is shown in Fig.\,2(d) along the $\mathrm{\overline{M} - \overline{X} - \overline{M}}$ direction. Red lines show the \DUV calculation, which nicely reproduces not only the surface state dispersion, but also the dispersion of the VB bands at lower energies. 
Upon optical excitation, in Fig.\,2(e), electrons are excited well above $\mathrm{E_F}$, and the band dispersion appears kinked. The band velocity is reduced with respect to the equilibrium one, and it is now better described by the simple DFT calculation shown as yellow lines. This effect will be analyzed quantitatively in Fig.\,3. Here we notice that the timescale of the band renormalization is within the temporal resolution of our setup. Already 300 fs after the arrival of the pump pulse the band dispersion has recovered the equilibrium slope [Fig.\,2(f)], only with the electronic system at a larger effective temperature. The ultrafast timescale suggests a purely electronic origin for the light-induced band renormalization. 
In particular, the observed kink in the optically excited band structure cannot be ascribed to the coupling to a phonon, which would affect the dispersion in limited energy windows both above and below $\mathrm{E_F}$ corresponding to the energy of the specific mode.  By contrast, the observed change in dispersion extends well above the largest phonon energy (50 meV) in ZrSiSe \cite{Zhou_PRB_17,Duman_16}.


\begin{figure*}[]
  \centering   \includegraphics[width = 0.95 \textwidth]{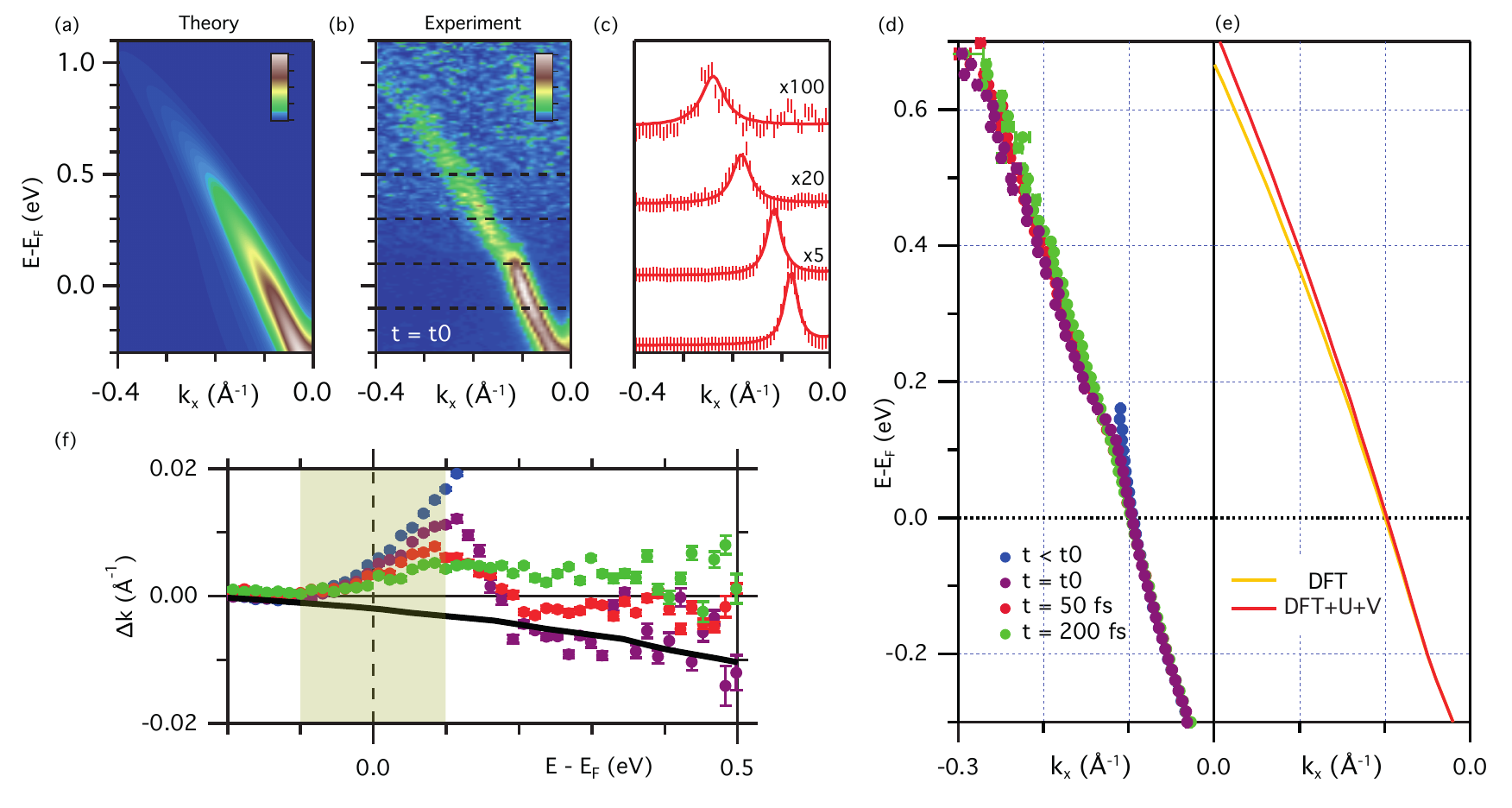}
  \caption[didascalia]{(a) Simulated ARPES intensity from the \textit{ab initio} calculations for U = 1.47 eV and V = 0.33 eV, broadened by the experimental momentum and energy resolution. The electron occupancy is defined by a Fermi-Dirac distribution for an effective temperature T* =  600 K, as inferred from the experimental data at 200 fs. (b) Experimental spectral function at t$_0$, during optical excitation. (c) MDCs at selected energies, indicated in (b) by dashed lines, and Lorentzian fits. (d) Peak positions as a function of energy; colors correspond to different delay times. (e) Calculated band dispersion for \DUV (red) and DFT (yellow), respectively.  (f) Change in band position, $\Delta$k, obtained as the difference between the experimental and peak position calculated with  \DUV. The black line indicates the theoretical $\Delta$k obtained as difference between the DFT and the  \DUV calculations.

   }
  \label{fig:tr_ARPES}
\end{figure*}


Theory and experiment are quantitatively compared in Fig.\,3. We simulate the ARPES intensity in Fig.\,3(a) starting from the \DUV  calculation, broadened to account for the experimental resolution. The band occupancy is determined by the effective electronic temperature (T*) estimated from a Fermi-Dirac fit to the experimental data 200 fs after optical excitation (T* $\sim$ 600 K) \cite{Supp_ZrSiSe}. The experimental dispersion is determined from Lorentzian fits to the momentum distribution curves (MDCs) extracted from the data, as shown in Fig.\,3(c) for selected energies. The QP energies are shown in Fig.\,3(d), where different colors encode the corresponding delay times during and immediately after optical excitation.  The flattening of the band is well captured by the change in dispersion between the \DUV (red) and DFT (yellow) calculated bands, as shown in Fig.\,3(e). We plot in Fig.\,3(f) the momentum renormalization $\Delta k$, i.e. the difference between the experimental (markers) and DFT (black line) dispersion with respect to the dispersion calculated by \DUV. 
For $E - \mathrm{E_F}\geq 150$\,meV $\Delta k$ is negative and as large as $- 0.01\pm 0.003$\,\AA$^{-1}$. The renormalization is largest during optical excitation (purple markers), and suddenly decreases after the pump pulse (red markers). The momentum renormalization $\Delta k$ is found to vary with the pump fluence, as discussed in the supplemental material \cite{Supp_ZrSiSe}.

Finally, we address the band dispersion near $\mathrm{E_F}$ in the shaded area in Fig.\,3(f). The positive $\Delta k$ observed before the arrival of the pump pulse is a trivial apparent deviation from the band dispersion due to the sharp Fermi-Dirac cutoff. This effect is well-known in ARPES and is stronger at low temperature  \cite{Hofmann_APA_05}. Interestingly, $\Delta k$ remains positive during optical excitation, which indicates that electrons have not yet reached a thermal distribution. According to our analysis of the temporal dynamics of the Fermi-Dirac distribution \cite{Supp_ZrSiSe}, for the fluence used in the data set of Fig.\,3, electrons fully thermalize via electron-electron scattering after 200 fs (green markers). Only when a hot thermalized electron distribution is established does the broader Fermi distribution cancel the positive $\Delta k$, and the band follows the \DUV dispersion both around $\mathrm{E_F}$ and at larger energies. This observation establishes a clear hierarchy between the timescales of the band renormalization and of electron-electron scattering and thermalization in ZrSiSe.


In summary, by combining time-resolved ARPES and \textit{ab initio} \DUV calculations we have shown that correlations can be optically modified in ZrSiSe, resulting in a QP renormalization. The equilibrium band structure is well reproduced by moderate on-site $U$ = 1.47 eV and an inter-site $V$ = 0.33 eV Coulomb terms, consistent with the reduced electronic screening of the Dirac QP. Upon optical excitation, the enhanced screening by high-energy electrons and holes produces a measurable change in the band dispersion.
Previous tr-ARPES experiments have revealed ultrafast changes in the band dispersion of strongly correlated electron systems such as the high-temperature cuprate superconductors (HTSCs). They show changes of the coherent spectral weight \cite{Smallwood_Sci_12}, of the QP scattering rate \cite{Boschini_NatMat_18} or of the QP effective mass \cite{Rameau_PRB_14}.  These effects have been interpreted in terms of an optically induced reduction of the phase coherence, a mechanism which is specific to HTSC. Here we have shown that the QP dispersion can be controlled by purely electronic means, by changing the electronic screening of the Coulomb interaction. Our results are the first experimental evidence of a more general mechanism, originally proposed to control the electronic dispersion of correlated metal oxides \cite{Nicolas_PRL_12}, which our study extends to the Dirac QP in NLSMs. Our findings demonstrate how ultrafast optical doping can be used as an alternative way of controlling quasi-particle properties, by tuning many-body interactions on a fs timescale. In particular, this could be exploited in other topological materials with linearly dispersing Dirac and Weyl states, where electronic correlations are believed to enhance the electron mobility \cite{Fujioka_NatCom_19} or to induce Lifshitz transitions \cite{Xu_PRL_18}.



We acknowledge  financial support by the Swiss National Science Foundation (SNSF), via the NCCR:MUST and the contracts No. 206021-157773, and 407040-154056 (PNR 70).
This work was supported by the ERC Advanced Grant H2020 ERCEA 695197 DYNAMOX, the ERC-2015-AdG694097, the Cluster of Excellence (AIM), Grupos Consolidados (IT1249-19) and SFB925. The Flatiron Institute is a division of the Simons Foundation.
S.M. acknowledges support by the Swiss National Science Foundation (Grant No. P300P2-171221). This research used resources of the Advanced Light Source, which is a DOE Office of Science User Facility under contract no. DE-AC02-05CH11231.






\end{document}